%% file: article_acm.tex
\documentclass[sigconf]{acmart}

\usepackage{amsmath,amsfonts}
\usepackage{algorithmic}
\usepackage{graphicx}
\usepackage{textcomp}
\usepackage{float}

\usepackage{url}
\usepackage{framed}
\usepackage{tabularx}
\usepackage{booktabs}
\usepackage{comment}
\usepackage{float} 
\usepackage{colortbl}

\usepackage{changepage}
\usepackage{tcolorbox}
\usepackage{enumitem}

\usepackage[super]{nth}

\def\BibTeX{{\rm B\kern-.05em{\sc i\kern-.025em b}\kern-.08em
    T\kern-.1667em\lower.7ex\hbox{E}\kern-.125emX}}

\usepackage[framemethod=TikZ]{mdframed}
\definecolor{backgroundpage}{HTML}{F8F5F4}

\newenvironment{coloredframe}[2][]{
    \mdfsetup{
        skipabove=3pt, skipbelow=3pt,
        hidealllines=true, leftline=true,
        innerlinewidth=1pt, innerlinecolor=#2, 
        linewidth=0pt,
        backgroundcolor=#2!10
    }
    \begin{mdframed}}
    {\end{mdframed}}

\newcommand{\quotehans}[2]{
   \begin{coloredframe}{yellow}
    \textit{``#1''} #2.
    \end{coloredframe}
}

\newcommand{\quotehanna}[2]{
    \begin{coloredframe}{teal}
    \textit{``#1''} #2.
    \end{coloredframe}
}

\definecolor{interviewcolor}{HTML}{b8b0ff}

\newcommand{\interviewquote}[2]{
   \begin{coloredframe}{interviewcolor}
    \textit{``#1''} #2.
    \end{coloredframe}
}

\AtBeginDocument{%
  \providecommand\BibTeX{{%
    \normalfont B\kern-0.5em{\scshape i\kern-0.25em b}\kern-0.8em\TeX}}}

\copyrightyear{2024}
\acmYear{2024}
\setcopyright{rightsretained}
\acmConference[ICSE-SEET '24]{46th International Conference on Software Engineering: Software Engineering Education and Training}{April 14--20, 2024}{Lisbon, Portugal}
\acmBooktitle{46th International Conference on Software Engineering: Software Engineering Education and Training (ICSE-SEET '24), April 14--20, 2024, Lisbon, Portugal}\acmDOI{10.1145/3639474.3640072}
\acmISBN{979-8-4007-0498-7/24/04}

\begin{document}

\title{Breaking Barriers: Investigating the Sense of Belonging Among Women and Non-Binary Students in Software Engineering}

\author{Lina Boman$^*$, Jonatan Andersson$^*$}
 \affiliation{%
  \institution{University of Gothenburg \\
    Dept. of Computer Science and Engineering}
   \city{Gothenburg}
   \country{Sweden}
 }
 \email{{gusbomli,gusganlgjo}@student.gu.se}

 \author{Francisco Gomes de Oliveira Neto}
 \orcid{0000-0001-9226-5417}
 \affiliation{%
   \institution{Chalmers and the University of Gothenburg \\
    Dept. of Computer Science and Engineering}
   \city{Gothenburg}
   \country{Sweden}
 }
 \email{francisco.gomes@cse.gu.se}

\renewcommand{\shortauthors}{Boman, Andersson and de Oliveira Neto}

\input{sections/00_abstract.tex}



\maketitle

\def\thefootnote{*}\footnotetext{These authors contributed equally to this work}\def\thefootnote{\arabic{footnote}}

\input{sections/01_introduction}
\input{sections/02_relatedwork}
\input{sections/03_methodology}

\input{sections/04_results.tex}
\input{sections/05_discussion.tex}

\input{sections/06_conclusions.tex}

\bibliographystyle{ACM-Reference-Format}
\bibliography{bibliography}

\end{document}

%% file: sections/00_abstract.tex
\begin{abstract}
Women in computing were among the first programmers in the early \nth{20} century and were substantial contributors to the industry. Today, men dominate the software engineering industry. Research and data show that women are far less likely to pursue a career in this industry, and those that do are less likely than men to stay in it. Reasons for women and other underrepresented minorities to leave the industry are a lack of opportunities for growth and advancement, unfair treatment and workplace culture. This research explores how the potential to cultivate or uphold an industry unfavourable to women and non-binary individuals manifests in software engineering education at the university level. For this purpose, the study includes surveys and interviews. We use gender name perception as a survey instrument, and the results show small differences in perceptions of software engineering students based on their gender. Particularly, the survey respondents anchor the values of the male software engineer (Hans) to a variety of technical and non-technical skills, while the same description for a female software engineer (Hanna) is anchored mainly by her managerial skills. With interviews with women and non-binary students, we gain insight on the main barriers to their sense of ambient belonging. The collected data shows that some known barriers from the literature such as tokenism, and stereotype threat, do still exist. However, we find positive factors such as role models and encouragement that strengthen the sense of belonging among these students.
\end{abstract}

%% file: sections/01_introduction.tex
\section{Introduction}
\label{section:introduction}
The United Nations states the areas of Science, Technology, Engineering and Mathematics (STEM) field as one of the most unequal fields~\cite{united_nations2022} and software engineering is a significant part of the field. The reality that only one gender dominates the software engineering industry brings several issues. Several studies describe that companies without gender and ethnic diversity miss out on unique skills and perspectives~\cite{nunes2018innovation, hunt2015diversity, insights2011global,credit2016cs}, which hinders innovation and that such companies are more likely to perform worse financially~\cite{hunt2018delivering, peterson2016gender}. Diverse teams have an effect on performance as reported by a study comparing diverse and homogeneous teams solving a murder mystery~\cite{Phillips2009IsTP}. For instance, the diverse teams were better in solving the murder mystery, but at the same time believed that they were performing poorly. Further, looking at software engineering specifically, homogeneous teams produce worse quality software and make more errors than diverse ones~\cite{bosu2019diversity,catolino2019gender}.

The reasons behind the gender disparity in the software engineering field are multifaceted. One mainly contributing factor is the low number of non-binary or women who choose software engineering as a career. Further, among those that do, they are often not retained, and data shows that women are more likely to leave the field than their male peers at every level~\cite{ncwit_women_it_facts}. Research shows that women and underrepresented minorities are more likely to leave tech jobs due to a lack of opportunities for growth and advancement, as well as unfair treatment and workplace culture~\cite{womenwhotech2020, pew2018, mckinsey2022, allbright2020}. 

In workplace settings where a particular identity, such as gender, constitutes less than 30\% of the employees, a phenomenon known as tokenism threshold or critical mass theory can manifest~\cite{KanterRosabethMoss1993Mawo,RosetteAshleighShelby2010AWaC}. According to the theory, the individuals belonging to the minority group are more prone to being primarily perceived and evaluated based on their group identity rather than as unique individuals. In such instances, the low representation of the minority group compromises the minority group's potential to rise above stereotypical expectations and be acknowledged for their unique strengths. Another well-studied sociological phenomenon that women might experience in a male-dominated field such as software engineering is stereotype threat. Stereotype threat describes the anxiety or concern that an individual experiences when being aware of negative stereotypes associated with one's group and fear that their behaviour or performance may confirm those stereotypes~\cite{NguyenHannah-HanhD2008DSTA,SteeleClaudeM1995STat,spencer1999stereotype}. 

Other findings related to the subject have been found in women choosing or switching to non-STEM majors, which was researched by a study that looked at different levels of skills in verbal and math ability. Unsurprisingly, those that had high math and low verbal skills chose STEM majors, but those that had both, which most often were women, were more likely to choose non-STEM majors. One of the reasons for this was that women more often valued working with people~\cite{WangMing-Te2013NLoA}. This could mean that the software engineering industry miss out on those that are most attractive to it: skilled programmers with high interpersonal skills.


The significance of ambient belonging can be emphasised with several studies. It has shown to be a factor that affects students academic achievement~\cite{WaltonGregoryM2007AQoB, FroehlichLaura2023IoiS} and it shapes self-concept and understanding of self~\cite{MarkusHazelRose2010CaSA}. Having to face stereotypes is a also a factor that affects academic performance~\cite{obrien2003stereotype, SchmaderToni2008AIPM, NguyenHannah-HanhD2008DSTA}.

The purpose of this study is to examine the barriers that women and non-binary (W\&NB) individuals face that hinder their sense of ambient belonging in software engineering. More specifically, we explore the \textit{social factors} that play into ambient belonging. Although self-perception is a significant part of ambient belonging, this study focuses on social aspects, such as social interactions and culture. The study is not aimed to provide suggestions on improvements, but rather useful insight and knowledge. The focus of this study is on software engineering education at the university level, to gain insight into how tendencies for excluding cultures manifest. We conduct surveys to gather knowledge on how women are perceived among software engineering students. We also conduct nine interviews with W\&NB students with diverse backgrounds to gain insights into their experiences. 

Together, results from the surveys and interviews provide a nuanced image of ambient belonging among female and non-binary students in software engineering education. 
Ultimately, our findings provide insights that can be used for developing strategies to promote greater gender diversity in the field of software engineering in higher education. Our research also highlights known barriers and the importance of interventions and initiatives to cultivate an inclusive and supportive environment. Inspiration for such interventions and initiatives is found in what the participants' described as factors that positively contribute to their sense of belonging.

%% file: sections/02_relatedwork.tex
\section{Background and Related Work}
\label{section:related_work}
Understanding social science focused on minorities' experiences, social role theory, and how they manifest in other STEM fields is important to support the conclusions from our research. Therefore, we describe below the research context and the knowledge gap in where our study aims to fit. 

\subsection{Stereotypes and Sense of Belonging}
Social role theory can partly explain gender differences in STEM fields, as in one study that argues that it can be partially explained by gender roles and expectations that associate men with quantitative and scientific abilities, and women with interpersonal and verbal abilities~\cite{eagly2000social}. These stereotypes and expectations can influence individual's career choices and interests, as well as the way that they are perceived and treated by others in STEM fields. A clear example of this can be found in the tech industry, where many women experience that they are often pushed to managerial roles and rarely acknowledged for their technical abilities or interests~\cite{allbright2020}. This aligns with a the broader idea that individuals from underrepresented groups often experience pressure to conform to certain roles or behaviours in order to be accepted or valued in a particular domain or fashion. For women, this could be seen as their acceptance or recognition depending on conforming to stereotypical gender roles or assuming nurturing and care-taking responsibilities~\cite{foschi1996double,heilman2005same,eagly2002role}.

The influence of racial and gender stereotypes is reported in a study that looks at the value of acknowledgement and recognition from peers and educators~\cite{hughes_schellinger_roberts_2020}. This study also highlights that women are perceived as either competent or likeable, but rarely both, and that black girls and women are more likely to be punished for behaviours conflicting with the ``good girl'' student stereotype. The finding that women are perceived as either likeable or competent is also supported by the \emph{Hans and Hanna study} that describes how the same description of a person influences people's perception of them as a leader negatively if they have a female name instead of a male name~\cite{haurdic2018litar}. The basis for these findings were a survey, where the participants all received a text that had the same description of a leader, with the only difference being that name of the person they described was either Hans or Hanna. We apply the same strategy for our survey and found results that both contrasted and aligned with the existing research (details in Section \ref{sec:results}).

One study about ambient belonging in computer science looked specifically at how the environment in which students learn computer science affects their ambient belonging~\cite{cheryan2009ambient}. The different environments used in the experiment were either decorated with objects considered highly stereotypical of computer science or decorated with more neutral objects. The study found that women were less likely to find ambient belonging or be interested in the former environment and concluded that the stereotypes ascribed to computer science cause a lower interest in the field by women. Moreover, another factor to women's retention is the role of \textit{imposter syndrome}\footnote{The phenomena where one believes that their competence or skills are insufficient to be taking up space in the environment that they are in, often leading to the fear or anxiety that they will be called out for it.} that has been emphasised in several studies which have found that students who experience imposter syndrome may struggle with their sense of belonging and may question their abilities and fit within the STEM field and other male-dominated fields~\cite{clance1978imposter,perez2014role,sakulku2011impostor,stout2011steming}. Consequently, these feelings of self-doubt and inadequacy can contribute to decreased motivation, engagement, and persistence in their chosen STEM majors. Another study, a literature review, examined the impact of gendertyped cues, such as the presence of masculine interests and activities, on women's interest and participation in STEM fields~\cite{CheryanSapna2017WASS}. They proposed that the presence of gender-typed cues in STEM fields creates a masculine culture that is less welcoming and inclusive for women. In turn this can lead to a self-reinforcing cycle, where the lack of female representation in the field perpetuates the perception that the field is not for women. Another interesting finding in this study was that gender ratios in STEM fields are strongly related to the extent to which people in those fields believe that raw, innate talent is the key to success, so that fields such as physics had fewer women than fields in which people believe that hard work and persistence are the keys to success, such as biology.

The discrimination that those with intersecting marginalised identities face is explained by a theory called intersectionality~\cite{crenshaw1991mapping}. This can also be found in studies about STEM fields such as one that found that women of colour in STEM experience a ``double bind'' of race and gender-based discrimination, potentially leading to feelings of isolation, exclusion, and a lack of sense of belonging~\cite{ong2011inside}. They also found that women of colour in STEM often have to navigate negative stereotypes and assumptions about their abilities and competence, which can undermine their confidence and self-esteem. The unique barriers that those in STEM who have intersecting marginalised identities with gender and sexuality are shown in a study that found that LGBT professionals in STEM often experience challenges rooted in hetero-normative and gendered assumptions and biases that are deeply ingrained in STEM culture~\cite{cech2019lgbt}. These challenges were related to workplace climate, networking, mentoring, and career advancement. These studies highlight the need for greater awareness and understanding of the unique challenges faced by professionals with intersecting marginalised identities to better understand gender disparities in software engineering.

\subsection{Peer interactions and Representation}
Individuals who base their self-worth on external factors, such as the approval of their peers, are more vulnerable to negative outcomes such as academic disengagement, poor performance, and decreased well-being~\cite{crocker2003contingencies}. As female and non-binary students are more likely to face negative stereotypes and biases from their peers, these factors may be particularly relevant.

Another study that touches on how female students are perceived is one that studied students, teachers and parents in K-12, their perception and understanding of computer science as a research field, while also looking at differences between both gender and race~\cite{Hong2016}. One of their most relevant findings was that media shows a narrow perception of who does computer science, being mainly ``white, male and wearing glasses''. Girls were also less likely to respond that they regularly saw people like themselves in the media. Another finding was that Hispanic students were less likely to perceive themselves as skilled in the subjects that were perceived as necessary to do SE \& CS and had less confidence that they could learn it, than White or Black students. In contrast, the parents who encourage their children to pursue this field varied across gender and race. Still, there is a lack of exploratory research understanding the reasons behind these perceptions and why experiences vary so clearly among across different groups. Despite our small sample of participants, our interviews offer some useful insights.

%% file: sections/03_methodology.tex
\section{Research Methodology}
This study explores how social barriers that lead to female and non-binary individuals leaving the software engineering field manifest in software engineering education. Therefore, we focus on the following research questions (RQ): \\

\noindent \textbf{RQ1: How are female students in software engineering perceived by their peers?}

We aim to understand the differences in perceptions based on gender. Therefore, we apply a modified version of a survey called \emph{Do you trust Hans or Hanna}~\cite{haurdic2018litar} to students in a Software Engineering Bachelor (BSc) Program.\\

\noindent \textbf{RQ2: What social barriers do female and non-binary students in IT face that hinder their sense of ambient belonging?}

We aim to gather insights on the experiences of W\&NB students through semi-structured interviews. Interviews provide insightful knowledge as it allows researchers to listen and analyse the individual's experience (here, in the context of their education). \\

To ensure consistency throughout the interviews, (i) we design an interview guide to adhere to some structure, and (ii) one of the researchers conducted all the interviews. We anonymise the data to ensure privacy for the participants. In addition, the research follows the procedures for informed and freely given consent described in~\emph{GDPR Art.7}~\cite{gdprart7}. For example, this includes the participants being informed and understanding how the data is used and that the participant signs a consent form. Most importantly, the participants are informed that they are free to opt-out of the study at any time. 


\begin{figure}[ht]
    \includegraphics[width=0.8\linewidth]{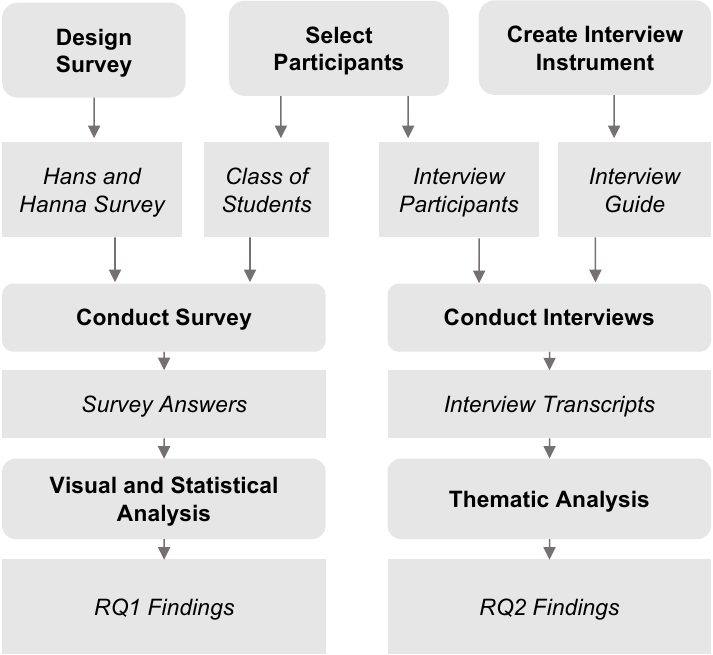}
    \centering
    \caption{Overview of our research method. The shapes distinguish between an activity done by researchers (rounded rectangles) and the corresponding artefacts such as documents, files, etc. (rectangles).}
    \label{fig:methodology}
\end{figure}

Next, we detail each step in our methodology (as shown in Figure~\ref{fig:methodology}). We use convenience sampling to send and collect survey answers by drawing  students from the Software Engineering and Management Program at the University of Gothenburg (Sweden). For the interviews we use a combination of convenience and purposive sampling where we intentionally select a diverse sample of W\&NB students within software engineering.\\

\noindent \textbf{Survey Design and Application:} The survey design was inspired by the Hans and Hanna study~\cite{haurdic2018litar}, which focuses on the perception of an industry leader. We adapted the survey to present a vague description of a software engineering student (Hans or Hanna), allowing participants' perceptions to fill in the gaps in the story. The story is written in a way so that there are no right or wrong answers to the questions of how skilled, competent, or likable the person is. 
The story used is more condensed than the original Hans and Hanna story to convey the fact that the person described is at the end of their studies and the very beginning of their career, meaning there is not as much to say that could affect the perception of their competence. Since this survey's main goal is to deduce perception based on gender, other traits are also reduced to ensure the subjects make assumptions solely based on gender. 

The survey was printed and shared with software engineering undergraduate students finishing their first-year of education (April 2023). A course instructor allowed us to use a part of their lecture to administer the survey in classroom. After the lecture was finished, we informed students of our ethical guidelines and, those that chose to participate, were informed to be split in two equal sized groups. We asked students to spread to different places to avoid friends sitting in the same group, or discussing the questions with each other. One group responds to the survey based on a description of a software engineer named Hanna, while the other group receives the same survey but with the name changed to Hans. We transcribe the survey data into a digital format for subsequent data analysis.\\

\textbf{Interviewing Protocol:} This interview is conducted with female and non-binary students due to the scope of RQ2. Some of the found challenges could also be a reality for men but previous research states that future research should focus on women's experiences and not compare it with the male experiences~\cite{bergstedt_2018}. Nonetheless, we aim to expand out analysis to include male students and other universities in our future work. Each interview lasted between 45--60 minutes and participants chose to attend in person or via video conferencing software. The interview guide, shared in our data package. Interviews are audio recorded with the participant's consent and transcribed. The data transcriptions are thoroughly reviewed to ensure the quality of the transcription. \\

\textbf{Data Availability:} We make the survey answers, the survey forms, interview guide, the scripts used to generate our figures and tables in our Zenodo package~\cite{replication2023belonging}. The transcripts of interviews include private information about our participants, hence we share mainly the quotes reported in the paper.

%% file: sections/04_results.tex
\begin{table*}[htbp!]
    \centering
    \caption{Survey questions and the median answers for each level (Hanna or Hans). The rows with questions with a median difference are highlighted. The levels were: (1) Strongly disagree, (2) Disagree, (3) Neutral, (4) Agree, (5) Strongly agree.}
    \label{tab:question_overview}
    \begin{tabularx}{\linewidth}{llXrr}
    \toprule
\rowcolor{lightgray!25} \textbf{ID} & \textbf{Category} & \textbf{Text} & \textbf{Hanna} & \textbf{Hans} \\
\midrule
--- & Demographics & What are your pronouns? &   &  \\
Q01 & Competence & I trust in X's competence as a software engineer. & 4 & 4 \\
Q02 & Competence & I would ask X for help if I had any issues with my computer or other technical problems. & 4 & 4 \\
Q03 & Competence & I believe X's competence could be beneficial for my own group projects. & 4 & 4 \\
\rowcolor{blue!10} Q04 & Trust & I believe X could be a good scrum master. & 4 & 3 \\
Q05 & Competence & I would value X's opinion in a difficult technical decision. & 4 & 4 \\
\rowcolor{blue!10} Q06 & Trust & I believe X could be a good product owner. & 3 & 4 \\
Q07 & Competence & I believe X could be a good software architect. & 4 & 4 \\
Q08 & Competence & I would trust X in qualitative tasks such as Code reviews or manual testing. & 4 & 4 \\
Q09 & Open-ended & Can you share some details on some of your answers from the questions above? & --- & --- \\
Q10 & Competence & I believe X's experiences could be beneficial for my own group’s projects. & 4 & 4 \\
\rowcolor{blue!10} Q11 & Likeability & I believe that me and X have a lot in common. & 3 & 4 \\
Q12 & Likeability & I would like to work with X. & 4 & 4 \\
\rowcolor{blue!10} Q13 & Likeability & I would like to grab a beer/drink/coffee with X. & 3 & 4 \\
Q14 & Likeability & I would recommend X to a future employer. & 4 & 4 \\
Q15 & Trust & I would trust X in a leading position in my team. & 4 & 4 \\
Q16 & Likeability & I think X is qualified for them position. & 4 & 4 \\
\rowcolor{blue!10} Q17 & Likeability & I feel I can relate to X. & 4 & 3 \\    
Q18 & Open-ended & Optional: Describe what you find (or don't find) relatable about X: & --- & --- \\
\bottomrule
\end{tabularx}
\end{table*}

\section{Results}
\label{sec:results}

This section presents the results from our data collection and answers to our research questions. The quantitative and qualitative survey results aim to answer RQ1, whereas the qualitative interview results aim to answer RQ2.

\subsection{RQ1: Survey Results}

A total of 42 participants answered the survey split equally between both \textbf{groups}: Hans and Hanna. We asked participants their preferred gender pronouns: (i) she\slash her, (ii) he\slash him, (iii) they\slash them, (iv) other, (v) prefer not to say. Both versions of the survey had 90\% participants choosing the he\slash him pronouns. We divide the survey questions into three different themes, namely: trust, competence and likeability. Table~\ref{tab:question_overview} shows the questions as well as the median answers from each group.

Overall, there is no stark difference between the questions, such that most participants either agree, or are neutral for all questions. There is no difference between all questions related to Hanna\slash Hans's perceived competence. Albeit small, there are a few differences between respondents regarding questions Q04, Q06, Q11, Q13, Q17. Those questions cover trust and likeability. Regarding \textbf{trust}, we see a median agreement (median = 4) to the statement that Hanna could be a good scrum master, while being neutral to the statement that Hanna could be a good product owner (median = 3). The results for Hans are reversed.

We also see a median agreement that respondents have a lot in common with Hans (Q11), and would like to spend time with him (Q13). This is not surprising as 90\% (16) of our respondents identify as male. In contrast, there is a median agreement that respondents feel that they can \textit{relate to Hanna}. We refrain from drawing deeper conclusions from the results as both the sample size and median differences are small. Nonetheless, the median responses align with reported studies that see women in more managerial roles (Q04).

\begin{figure*}
    \centering
    \includegraphics[width=\linewidth]{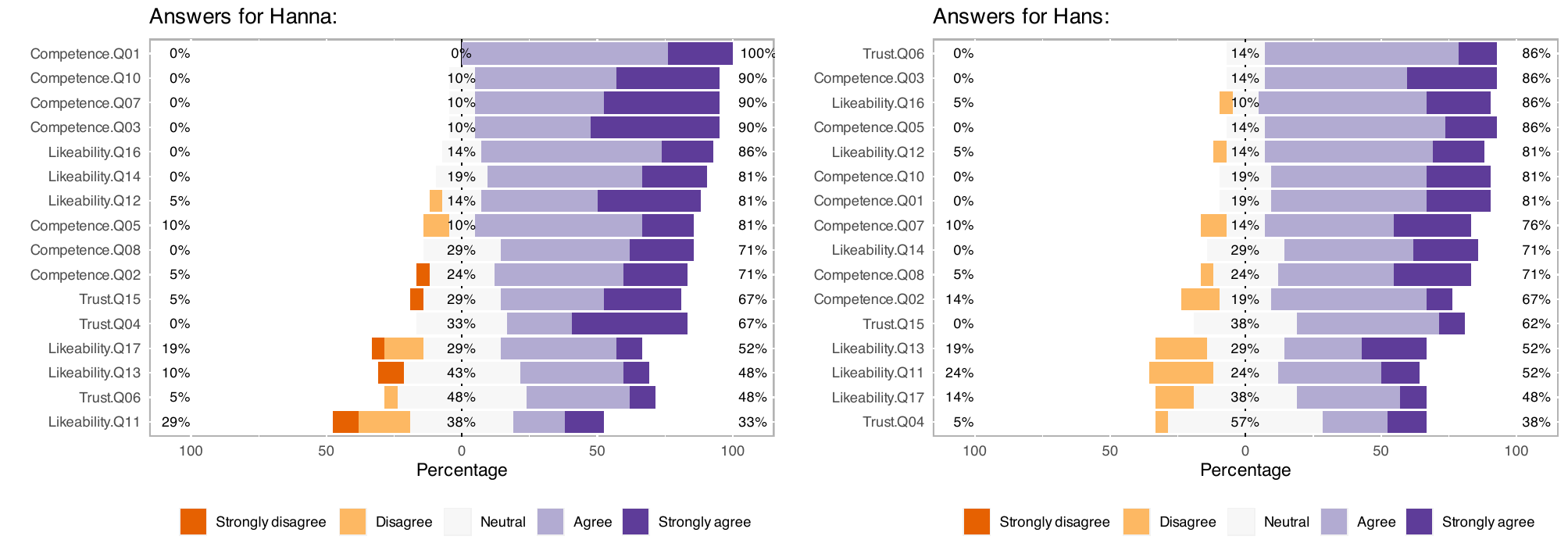}
    \caption{Diverging plot for the survey answers from Hanna (left) and Hans (right). Note that the questions (y-axis) are ranked based on the proportion of answers (higher agreement are on the top), hence being different for each group.}
    \label{fig:div_plot}
\end{figure*}

We look closer at the variance in the survey responses in a diverging plots for each group (Figure \ref{fig:div_plot}). First, we notice that respondents felt more strongly about Hanna as she has more extreme answers (Strongly agree\slash Strongly disagree) than Hans who tends to receive more Agree and Neutral answers. There are small differences between the ranking of the questions related to their agreement (which aligns to our observations from Table~\ref{tab:question_overview}), except for Q6 (Trust) that shows more disagreements in the Hanna group (bottom 2) when compared to Hans (top-1 answer). Therefore, 86\% of respondents agree that Hans could be a good product owner, versus almost half (48\%) for Hanna. The remaining questions that showed median difference (Q4, Q11, Q13, Q17) are ranked on similar places for both Hans and Hanna. 

Therefore, quantitative data suggest that the general difference in likeability, competence, and trust is unessential between Hans and Hanna. However, how the way that this manifests in the open-ended answers (Q09, Q18) is interesting. We have slightly paraphrased some examples of answers below to make them more readable. Respondents used the phrasings \emph{programmer} and \emph{software engineer} interchangeably.

The general perception is positive for both Hans and Hanna. Interestingly, participants also strengthen the argument of Hanna's value being anchored to her ability to be detail-oriented, or that she could lead, or her managerial skills. When participants discuss what they anchor Hans's value as a software engineer, the answers are more multi-faceted, such as his skills in embedded systems or his excitement of programming.

\quotehanna{I felt that Hanna is a positive leader, therefore she can lead meetings. In addition, I think she would like to help students if someone needs it.}{SP9 (he\slash him), Hanna}

\quotehans{Since Hans favourite course was embedded systems, I would assume that he excelled at that course. Only a good programmer would truly know how software worked under the hood.}{SP8 (he\slash him), Hans}

\quotehans{Hans feels like a good general image of a software student. I can relate to his experience, his excitement about programming and active position in a group.}{SP17 (she\slash her), Hans}

Below, respondent SP21 (she\slash her) has an interesting reflection stating that Hans's personality is found intimidating. The respondent is especially using the term ``imposter syndrome'', this is one of the extracted themes from RQ2 and is commonly mentioned throughout the interviews conducted where only W\&NB students were interviewed.

\quotehans{Hans is relatable in his curiosity and interest in software engineering, software architecture and upbringing. For me however that is where the similarities end. I admire that Hans takes initiative and makes his ideas known but I have impostor syndrome because of people like Hans. That leads into a cycle of people like me holding back their ideas or leaving the field entirely, and people like Hans flourishing and being heard over others.}{SP21 (she\slash her), Hans}

Several respondents relate to Hans and Hanna having ``strong opinions'' or their managerial skills. Something to note is that when it comes to comments about respondents' ability to relate to Hans, they are not solely based on his ``strong opinions'' or managerial skills, but rather multiple different aspects of him, such as balance or having strong technical skills.

\quotehanna{I agree with having strong opinions about how things should be done. I also like to take a leading role in projects. I am also organized and active in discussions.}{SP2 (he\slash him), Hanna}

\quotehanna{What I found relatable was her passion, however I do not have strong opinions on how this should be done. I try to be more adaptable to the situation.}{SP6 (he\slash him), Hanna}

\quotehans{He is a man of balance. I like that he makes time for what he likes to do and when he does work he does it efficiently and strongly. I would definitely trust a guy who not only understands the computer, but also people.}{SP5 (he\slash him), Hans}

One other aspect of the open-ended questions is the explicit use of the word ``trust''. For the Hanna survey, trust is never used in a positive sense. However, for the Hans survey trust is positively mentioned several times. In the Hanna survey distrust can be interpreted from some participants.

\quotehans{The main issue is the phrase \emph{has strong opinions about how things should be done}. He seems like someone not very flexible, yet I still think he is trustworthy.}{SP15 (he\slash him), Hans}

\quotehanna{Hanna is still a student. She has no professional experience. If I would have to chosen a software engineer I'd choose based on experience}{SP3 (he\slash him), Hanna}

\quotehanna{From the description she sounds like an enthusiastic learner and is doing rather well in her courses, so I would be happy to work with her. But that alone does not demonstrate her ability or level of competence.}{SP15 (he\slash him), Hanna}








\begin{tcolorbox}
\textbf{RQ1: How are female students in software engineering perceived by their peers?}\\ \\
The data from the surveys indicate that, on average, there are \textit{no major differences between the perceptions of genders} in term of how competent, trustworthy or likable they are. However there is a bigger variance in the perception of females. Terms such as ``trustworthy'' is only mentioned in the description of a male student. Female students competence are more likely to be evaluated by their managerial skills, while male students are evaluated on multiple skills.
\end{tcolorbox}

\subsection{RQ2: Interview results}
\label{sec:interview_results}

We conduct nine interviews, with eight participants identifying as a woman and one participant identifying somewhere on the woman to non-binary scale. A short description of the different interview participants is found in table, see Table ~\ref{tab:interview_subjects}. The participants are collected from two different universities, four different years of education and with an age disparity of 19--33. 

\begin{table}[ht]
    \centering
    \caption{Interview subjects presentation. We assign an ID to each participant.}
    \label{tab:interview_subjects}
    \begin{tabularx}{\linewidth}{llX}
        \toprule
        \textbf{ID} & \textbf{Age} & \textbf{Description}\\
        \midrule
        P1 &23 & Mainly interested in user interaction\\
        P2 &24 & Mainly interested in product owner related tasks \\
        P3 &26 & Interested in the problem solving aspects of software engineering \\
        P4 &20 & Enjoys programming \\
        P5 &25 & Enjoys back-end development, current goal is to become a Scrum Master \\
        P6 &19 & Main interest is software architecture \\
        P7 &29 & Wants to focus on back-end development \\
        P8 &25 & Master student focusing on business possibilities with software solutions \\
        P9 &33 & Current goal within software engineering is to develop better cloud computing skills\\
        \bottomrule
    \end{tabularx}
\end{table}

\begin{figure}
    \centering
    \includegraphics[width=\linewidth]{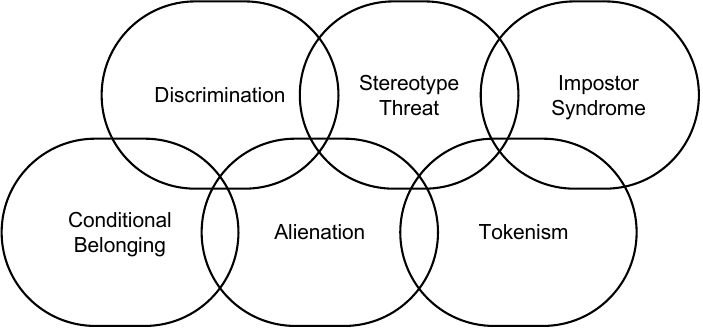}
    \caption{Thematic framework over the found social barriers to sense of ambient belonging in software engineering.}
    \label{fig:thematic_framework}
\end{figure}

We analyse the interviews through open-coding and thematic analysis. To align similar coding two researchers coded one interview independently and reviewed the results together and discussed the differences. From the extracted codes six themes emerged, described in Table~\ref{tab:main_themes}. Many experiences reported by our participants belong to the intersection of different themes, which is why the we conceptualised our themes as a Venn diagram (Figure~\ref{fig:thematic_framework}.

\begin{table*}[ht]
    \centering
    \caption{Themes within ``social barriers'' to ``sense of ambient belonging'' in Software Engineering education.}
    \label{tab:main_themes}
    \begin{tabularx}{\linewidth}{llX}
        \toprule
        \textbf{ID} & \textbf{Theme} & \textbf{Description}\\
        \midrule
        T1 & Stereotype threat & The fear of confirming negative stereotypes associated with one's underrepresented group. \\
        T2 & Imposter syndrome & The belief that one's skills are not as developed as they should be to belong to the current environment. \\
        T3 & Tokenism & When one feels that they are being evaluated based on the underrepresented group that they belong to, rather than as an individual. \\
        T4 & Alienation & The feeling of not belonging or not fitting in the environment or a context. \\
        T5 & Conditional belonging & Describes a belonging or acceptance that is based on conformity to certain roles or behaviors.\\
        T6 & Discrimination & Unjust treatment based on certain characteristics.\\
    \bottomrule
    \end{tabularx}
\end{table*}

\textit{All} the participants mentioned how, when applying to their programs, they perceived the stereotype of a software engineer to have at least some truth, and some of them considered other programs due to the fact that they did not feel like they could relate to that stereotype. For instance, Participant P8, also a master's student, even described that she chose a different bachelor's program due to the stereotype.

\interviewquote{I think I knew about the stereotypes before applying to [University], and that was probably one of the reasons why I didn't choose the software bachelor. Because I was like, I'm not fitting in.}{P8}

Several participants described examples of how \textbf{stereotype threat} is a barrier. Participant P5  recalls a situations where she felt that she was not trusted, and further described how she often for that reason would mention that her father is a tech manager. 

\interviewquote{I don't know if it's because of my gender or that I have an inkling that it does. So the way that I would kind of counteract that would be by triangulating someone who's established and male...}{P5}

Another participant who identifies as queer, expressed a need to tone down certain aspects of her when asked about how they think they are perceived gender-wise. The quotes below from the same participant conveys how they experience stereotype threat:
 
\interviewquote{Yeah, I tried to be a little bit passing... to avoid the conflict or potential of it.}{P6}

\interviewquote{I feel like it's more important that I made a mistake, for example. More than my male counterpart who made a mistake. Like there's much more pressure to be better.}{P6}

A clear majority of the participants mentioned struggling with \textbf{imposter syndrome}, particularly when working with other people in, e.g., group projects.

\interviewquote{I kind of had imposter syndrome where I was like, am I good enough? Am I stupid?}{P5}

\interviewquote{People are, I don't know, I feel like other people know so much more than me. Like, I'm unsure if I'm going to make it or not, because of my skills.}{P8}

There are experiences from participants' describing how the participants are seen as their gender, race, or sexuality instead of as individuals. This is a phenomenon called \textbf{tokenism threshold}. This barrier is referred to here as ``tokenism". One example of this is described in the context of being over-supportive and how it could have an opposite effect on some individuals. Participant P1 shares that she feels seen as a simplified woman, and not so much as a multi-dimensional individual.

\interviewquote{But when it's very supportive like, `oh, no, yeah, you're a woman in IT!'  ...I don't know, that kind of puts me off a bit...\\...sometimes it can be almost like having the opposite effect, because it makes me feel kind of smaller...\\...you wouldn't say that to a guy, I guess. Like, `you're a guy in IT!' or something.}{P1}

We illustrate the overlap between stereotype threat and tokenism with a quote where Participant P5 shares an incident where a teammate in a group project expressed themselves inappropriately. Similarly, Participant P6 felt that it was very palpable that the amount of female students in their year is very low.

\interviewquote{It's so subtle that I'm just kind of like, if I feel like if I call it out, then I'm drawing attention to it. And then I look dramatic, so I just let it go, which is like, what can I really do? If I address it, then I stir up ruckus. And then we'd have to have a  conversation... Especially as the only girl in the team.}{P5}

\interviewquote{I don't know, it kind of feels  we're competing with each other, I guess. We're competing to stand out.}{P6}

Participants describe \textbf{alienation} in different ways. For instance, P5 shares their struggles to find role models among teachers. In turn, P1 shares about the first year of her program, which was during COVID. She explains that the reason for her feeling this way is due to not being able to meet other female students to connect with.

\interviewquote{I don't think I see anyone as a role model. Yeah, especially since we don't have a lot of female teachers.}{P5}

\interviewquote{I didn't really meet a lot of people that had similar interests or something like me that I felt like I could relate to, and that made me think \emph{maybe I don't really belong here} and stuff like that, I think.}{P1}

The quote below illustrates the intersection between alienation and stereotype threat, where P1 reinforces the feeling of being a minority.

\interviewquote{...in the classroom and stuff like there's a lot of guys and they're always like asking questions. Especially in the first year, I thought that was very... I guess that has a bit to do with gender... that it was kind of male dominated.}{P1}

Next, we illustrate the reported \textbf{discrimination} experiences described by our participants. Participant P3 shares her experiences in group projects, where her teammates did not let her help them, even if they are stuck. Considering intersectionality, it could be worth mentioning that this participant is a woman of color with religious expressions. 

\interviewquote{Again, the underestimated part. They think because you are female, you don't know much and stuff like that. When we help them, they kind of feel like... Sometimes they don't believe you, actually... You know, they don't trust your solution, until they get confirmation from someone else. And then I'm really like, `what the [heck], bro?'}{P3}

Similarly, Participant P2 recalls a situation where a male former teammate of hers had went from working in a mixed group, to an all male group. In parallel, Participant P9 reports on some dealing with negative stereotypes regarding her gender and sexuality.

\interviewquote{And I got into an all female group and he got into an all male group and he texted me: \textit{Oh, great. Things are gonna move a lot faster now}.}{P2}

\interviewquote{I've been gay since I was 15, and I can put boundaries, so... if someone says something, I just tell them like, `no, you're wrong', and I just don't forget, I mean, I don't go into conflict... it doesn't stick with me, I explain, and even then, if they still don't understand,  I'm like, `okay, you do you, if you want to be ignorant, be ignorant, but that's your issue, not my issue', right?}{P9}


Lastly, we saw \textbf{conditional belonging} emerging in some of our themes. For instance, P6 describe that they feel the need to conform in a certain way to find belonging.

\interviewquote{You know how men have their ``bro and bro'' and they trust each other easier, people who are the same kind of group together.  And I didn't really click as well with the girls in the class.  So it's kind of like. I'm trying to fit in. So as to not be left out, I guess.}{P6}

\begin{tcolorbox}
\textbf{RQ2: What social barriers do female and non-binary students in IT face that hinder their sense of ambient belonging?}\\

For women and non binary student we detected six main barriers to their sense of ambient belonging in software engineering. Conditional belonging, discrimination, alienation, stereotype threat, tokenism and imposter syndrome. The most prevalent barrier in the data was imposter syndrome and alienation. 
\end{tcolorbox}

%% file: sections/05_discussion.tex
\section{Discussion}

From the survey answers (RQ1), we argue that the first-year students of the Software Engineering undergraduate program at the University of Gothenburg do not see a significant gap between gender perceptions of software engineers. These results contrast some of the literature that indicate a bigger gender gap regarding perceptions of students and professionals in STEM. Here, we discuss a few possible reasons relating to the context of the Software Engineering Program in which we conducted our study, but causation can only be claimed with further research where the amount of participants is increased, or broadening the scope to multiple universities. 

We also acknowledge that the results might be due to our sampling bias, as answers were gathered at the end of a university course, where the teacher has consistently cultivated an inclusive environment through the course by using inclusive techniques and language. This active inclusion is even mentioned in the interviews.

\interviewquote{I think it was [Teacher's Name], they mentioned, when you code, try to like, involve everybody, if you want to choose gender, just don't put female and male, like, think of those things, and since I'm bisexual, in the LGBTQ part, I was like, yeah, I mean, I hadn't even thought about it, that it's something that we actually need to think about, and that is something that I don't think a lot of people catch up to fast, if you're not in the LGBTQ community}{P9}

\interviewquote{I appreciate when teachers mentioned like non-binary people and stuff. It does make me feel like, `oh,  there is a place for me.'}{P5}

Another possible explanation for the surprising results is that there is a shift in the mindset amongst students where a more inclusive attitude has converged to a norm. This is somewhat supported by the interview data where several interview participants mentioned the surprisingly positive experience of their peers and their sense of ambient belonging.

\interviewquote{Because I was technically pursuing something else before I applied, and it didn't work out. And I'm glad it didn't work out because I much prefer being here. So I think, to some extent, faith, brought me to this program and I do enjoy it much more than I would have enjoyed the other option. I think, yeah, I should be here.}{P6}

\interviewquote{...there is a sense of trust that my teammates have in me that they would let me have a kind of more leadership position}{P5}

As the purpose of RQ1 is to understand the perception of female and non-binary individuals by their peers, the results are strengthened throughout the data collected from the interviews. The general opinion from the interview participants is that their peers are more aware and inclusive than what literature described in Section~\ref{section:related_work} indicates, just as the data from the survey questionnaire implies. Despite research and data showing that women are rarely seen as competent while also seen as likable~\cite{haurdic2018litar}, our data from the research shows the opposite, where all answers from the competence-related questions had equal median for both Hanna and Hans. One noteworthy takeaway from the survey's open-ended question, regarding likeability, is that the participants only found one aspect of Hanna to be relatable and likable, while Hans is perceived to have \textit{several} relatable aspects.

Our interview results confirm the existing social barriers that female and non-binary students face to their sense of ambient belonging in software engineering. From the analysis of the thematic framework and the participants' responses, it is evident that stereotype threat, imposter syndrome, tokenism, alienation, conditional belonging, and discrimination are in fact experienced by these individuals. \textit{What should be acknowledged is that female and non-binary students' experiences are not monotonous}. With gender, other intersecting identities, such as sexuality, gender variance, ethnicity, race, and religious expressions, can further augment barriers that these individuals face. Looking at the results through a lens of intersectionality provides a nuanced understanding of challenges that highlights the need to apply inclusive strategies that address the multidimensionality of diversity. 

The most common theme among the participants was imposter syndrome. Many expressed that it is related to being a minority in class and therefore questioning whether they are a good fit. This barrier, as well as the other barriers, can be addressed with interventions and initiatives with the purpose to cultivate an inclusive and supportive environment. Some suggestions or inspiration to how the barriers can be addressed can be found in the factors that helped with the interview participants' sense of ambient belonging.

From the positive experiences reported in the interviews, a few recurring themes emerged that we refer to as \textit{positive factors} that help overcome the barriers mentioned above. Those themes were mentioned more often by participants that displayed more confidence in their sense of belonging to software engineering. The positive factors are described in Table ~\ref{tab:positive_forces} and cover the topics of (i) role models, (ii) encouragement, and (iii) community.

\begin{table}[ht]
    \centering
    \caption{Positive factors that decrease social barriers to the sense of belonging in software engineering education.}
    \label{tab:positive_forces}
    \begin{tabularx}{\linewidth}{lX}
        \toprule
        \textbf{Positive Factor} & \textbf{Theme}\\
        \midrule
        Role Models & Any individual that the participants can relate to that is an inspiration within software engineering. This could include family members, peers, influencers, or industry individuals. \\
        Encouragement & Positive feedback or inspiration to achieve a career within software engineering received by family members, peers, or role models\\
        Community & Describes a social group to share and experience the daily life as one pursues their career in software engineering.\\ 
        \bottomrule
    \end{tabularx}
\end{table}

\begin{figure}[ht]
    \centering
    \includegraphics[width=0.6\linewidth]{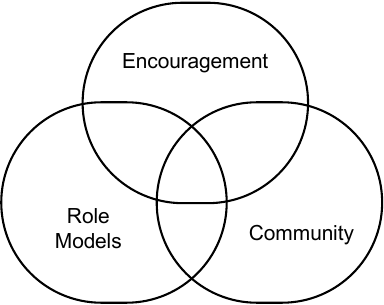}
    \caption{Thematic framework over Positive factors described by the interview participants}
    \label{fig:thematic_framework_positive_forces}
\end{figure}

To systematically discuss the positive factors we illustrate how the themes intersect in Figure ~\ref{fig:thematic_framework_positive_forces}. We also notice a relationship between the positive factors and corresponding social barriers. For instance, participants that sensed less ambient belonging are mainly held back by the barrier imposter syndrome.

A few interview participants describe several different experiences when they felt imposter syndrome. However, their imposter syndrome is no longer prevalent to the same extent. The following quotes convey how role models, encouragement and community work together to support W\&NB to find ambient belonging in software engineering.

\interviewquote{I do have (female) friends who are working in the field and they've been such huge support systems throughout my whole degree.}{P2}

\interviewquote{We are discussing that a lot in my friend group. So yeah, but I feel like I've overcome [imposter syndrome] now...\\I feel like now I can do anything. But in the beginning it was really hard...\\...the community part, and being a group of people being able to discuss these kinds of things helped me a lot to not be alone in this. So otherwise, if I wouldn't have friends in the same situation or other people to talk to, I would probably have quit.}{P8}

Participant P8 above also talked about a blog by a female software student that played a significant role in her interest to pursue software engineering.

\interviewquote{I was also reading a blog from [other anonymous University] in [other anonymous city]. They had a blog where students were writing about their daily life. I was following one girl who was studying software, actually. She was so inspiring. I was so close to choose a software bachelor's... \\ ...in the end, I chose industrial engineering management. 
}{P8}

We summarise our discussion in the points below as recommendations to educators, student organisations and Universities that aim for inclusive software engineering education. Even though those recommendations are limited by the scope of our research (e.g., sampling bias, choice of instruments, small sample size), we argue that the reported and lived experiences of our participants is a reflection of a positive change of direction, particularly as equity, diversity and inclusion claim more space in news outlet, social media and political discussion.

\begin{itemize}
    \item \textbf{Advocate for representation in teaching roles:} Our participants reported that they do not feel represented when looking at their teachers. While diversity among faculty is difficult to address, teachers and study administrations should aim for balance in the composition of supporting teaching roles, such as student representatives, or teaching assistants. These initiatives strengthen the factors of encouragement and role models, particularly for first-year students trying to build their sense of belonging in the study environment.

    \item \textbf{Encourage activities to connect students from minority groups:} Community was a recurrent theme among the positive forces, particularly in reducing the effect of impostor syndrome. Universities should encourage and allocate resources (e.g., time, or funds) to student unions or associations to organise activities for groups of students such as women or LGBTQIA+.
    
    \item \textbf{Promote role models:} Computer science history includes role models that belong to marginalised groups that are yet unknown to many students~\cite{santos2023scientists}. Bringing visibility and awareness to inspiring figures \textit{throughout their different courses} can help students from marginalised groups to find role models that they can relate and reinforce their sense of belonging in software engineering.
\end{itemize}

\subsection{Threats to validity}
\label{threats_to_vlidity}

Here, we address the different types of validity threats in our study that limit the scope of our research and can help future researchers in overcoming of exploring opportunities for further research. One of the construct validity threats is our choice of survey (describing Hans and Hanna) to represent the role of software engineers that students can relate to. Albeit a limitation, this choice helps with the reliability of our study as this type of study is proven to be repeatable and is used is in other fields to understand the differences in gender name perception~\cite{Gaustad_2015,haurdic2018litar}. Similarly, collecting data through interviews has disadvantages in terms of repeatability. Under the influence of factors such as interviewer bias, contextual factors, and participant interpretation, there may be variations in the data obtained. These disadvantages create issues concerning reliability and accuracy of the data. To address these issues, the interviews are semi-structured and follow an interview guide. 

Moreover, the researchers' own cultural biases and assumptions into the research process can have a similar effect as confirmation bias to hinder the conclusion validity of our study. Cultural biases threat the validity by misrepresenting or overlooking certain cultural perspectives or experiences related to the research topic. Another bias to be aware of is observer bias. This issue manifests if our physical presence or behaviour and the participants' potential awareness of our views or expectations lead to them altering their responses, to align with those expectations. The participants may also respond in a way that they perceive to be socially acceptable or desirable, rather than providing honest or accurate answers, introducing social desirability bias. The main methods used to mitigate this threat are the joint work by all authors in designing the survey instrument, interview guide and discussions aiming for agreement on themes during the thematic analysis.

One of the limitations in terms of scope is the focus on software engineering and not computer science as a whole. We argue that software engineering, compared to computer science, is also more broadly scoped than computer science as it is a mix of technical and non-technical subjects, with some learning goals focused on management, development processes, etc. The broadness of the scope allows for students to gain varied experiences, thereby participants being able to provide nuanced insight into how their experiences might contrast between subjects.  

In the survey questionnaire, data sampling, and results, we acknowledge the following threats to internal validity. The sample size is too small to be statistically significant. The sample is not randomly selected from different universities. 

In our survey, we recognize several threats to internal validity, including a small, non-random sample from a single university program, potentially influenced by shared social norms, teachers, and peers. The survey's inferential results, based on data patterns, align with interview outcomes, enhancing internal validity. However, the interview process, employing purposive and convenience sampling, involved participants familiar with the researchers, possibly affecting the results. An interview guide helped maintain consistency. Future research will expand to include diverse student cohorts from various universities internationally.


Lastly, the main external validity to this research is that, due to convenience sampling, most of the data is collected through our local university studying the same program. Therefore, we cannot claim that our findings can be generalised or transferred to different study environments. However, the investigated Program is international and attracts students from different areas of the world which helps avoiding a homogeneous sample of Swedish students and supports the repeatability on samples in other populations.

%% file: sections/06_conclusions.tex
\section{Conclusion}
Our study examines the factors causing women and non-binary (W\&NB) individuals to leave the software engineering field, particularly in educational settings. We pinpoint six major barriers faced by W\&NB in software engineering education and note that female students are often evaluated by their peers on fewer traits and competencies compared to their male counterparts. Additionally, we suggest strategies like fostering encouragement, building communities, and promoting role models to create a sense of ambient belonging for W\&NB in software engineering. These findings offer a foundation for future research aimed at improving W\&NB retention in this field.

We aim to explore multiples areas of future work, such as collecting more data on the experiences and perceptions of students concerning stereotypes, as well as doing new iteration of the Hans and Hanna survey for incoming students. This type of survey should try to focus on getting a larger sample to get better insights into the patterns found in this research and how they change over time. Lastly, we aim to do a longitudinal study focused on interventions to help mitigating those barriers.

\section*{Acknowledgements}

We extend our gratitude to all the respondents and participants who contributed to our study. Additionally, we acknowledge the support from the Gender Initiative for Excellence (GENIE) at Chalmers University of Technology,\footnote{\url{https://www.chalmers.se/en/about-chalmers/organisation-and-governance/equality/genie-gender-initiative-for-excellence/}} and the Lars Pareto grant from the University of Gothenburg.